\documentclass[12pt,prb,onecolumn,superscriptaddress,preprintnumbers,amsmath,amssymb,floatfix]{revtex4-2}

\usepackage{graphics,amssymb,amsmath,epsfig,color}
\usepackage{bm}

\usepackage{dcolumn}
\usepackage{bm}
\usepackage{float}
\usepackage{natbib}

\usepackage[export]{adjustbox}

\usepackage{braket}
\setcitestyle{square}

\usepackage{graphicx}
\usepackage{dcolumn}
\usepackage{bm}
\usepackage{float}
\usepackage{natbib}
\setcitestyle{square}

\usepackage{xcolor}

\usepackage{hyperref}
\hypersetup{colorlinks=true, linkcolor=blue, citecolor=blue, urlcolor=blue}

\usepackage{amsmath}
\usepackage{mathrsfs}

\DeclareFontFamily{U}{calligra}{}
\DeclareFontShape{U}{calligra}{m}{n}{<->callig15}{}

\begin{document}

\title{Ferroelectricity and chirality in  the Pb$_5$Ge$_3$O$_{11}$ crystal} %

\author{Mauro Fava}
\affiliation{Physique Th\'eorique des Mat\'eriaux, QMAT, CESAM, Universit\'e de Li\`ege, B-4000 Sart-Tilman, Belgium}

\author{William Lafargue-Dit-Hauret}
\affiliation{Physique Th\'eorique des Mat\'eriaux, QMAT, CESAM, Universit\'e de Li\`ege, B-4000 Sart-Tilman, Belgium}
\affiliation{Universite de Pau et des Pays de l’Adour, E2S UPPA, CNRS, IPREM, Pau, France}

\author{Aldo H. Romero}
\affiliation{Department of Physics and Astronomy, West Virginia University, Morgantown, WV 26505-6315, USA}

\author{Eric Bousquet}
\affiliation{Physique Th\'eorique des Mat\'eriaux, QMAT, CESAM, Universit\'e de Li\`ege, B-4000 Sart-Tilman, Belgium}

\date{\today}

\begin{abstract}
We study from first-principles calculations the ferroelectric structural phase transition of Pb$_5$Ge$_3$O$_{11}$ crystal.
The calculations of phonons and Born effective charges of the paraelectric phase allow us to identify a polar instability that is unstable in both transverse-optic and longitudinal-optic versions, giving rise to an entire branch of instability along a propagation vector parallel to the mode polarization (the hexagonal axe).
This is the hint of hyperferroelectricity and the stable head-to-head and tail-to-tail domain, as recently reported from both experiments and theory.
Then, our analysis of the ferroelectric phase shows that the polarization of Pb$_5$Ge$_3$O$_{11}$ is uniaxial along the hexagonal axes and with small in-plane components due to a piezoelectric effect.
The symmetry-adapted mode analysis shows that the total ferroelectric ground state distortion comes mainly from polar distortions of the unstable polar phonon mode but also from an invariant, cooperative mode that amplifies the polar deformation.
We also build a phenomenological model that highlights how the coupling between these modes is at play and helps us understand how to reproduce the second-order phase transition.
At last, we also quantify the structural chirality through the continuous symmetry measure method and trace its origin to the polar unstable mode itself.
By extending our approach to the phonon states we further show that the chirality is poorly affected by the relaxation but could also be enhanced by activating high frequency modes with polar symmetry.
Finally we study the phonon angular momentum (AM) distribution in both phases and identify trends in the AM behaviour across the Brillouin zone.
\end{abstract}

\maketitle

\section{Introduction}

The first synthesis of the compound lead germanate Pb$_5$Ge$_3$O$_{11}$ (PGO) dates back to 1971~\cite{iwasaki1971,nanamatsu1971}. 
Subsequent experimental works~\cite{Iwasaki1972} have established the main features of PGO, among which the existence of a high-temperature amorphous structure (T $>$ 618 K), an intermediate temperature hexagonal paraelectric (PE) phase  (450 K $<$T$<$ 618 K) with $P\Bar{6}$ symmetry and a low-temperature ferroelectric (FE) phase (T $<$ 450 K) with the trigonal $P3$ space group. 
The presence of a hysteresis loop for the natural optical activity,  i.e. gyroelectricity~\cite{aizu1964, wadhawan1979}, in the $P3$ ferroelectric phase, mirroring the one of the detected polar order, was also found in that early studies~\cite{Iwasaki1972,iwasaki1972b}, implying opposite handedness for the $+P$ and $-P$ state. 
Additionally, a sequence of experimental works on PGO have been focusing on the dielectric response, structural and ferroic properties~\cite{Iwasaki1972,newnham1973,kay1975,barsch1975,kirk1976,mansingh1979,venevtsev1981,shaldin2005}, on the piezoelectricity~\cite{yamada1972}, pyroelectricity~\cite{zwicker1976,bordovskii1978,takahashi1994,fadnavis1998,wazalwar2002} and the electro-optical properties~\cite{uchida1972,bichard1972,nordland1973,miller1974}.
Measurements of the spontaneous polarisation as a function of temperature reveal an uniaxial polarization along the hexagonal axis and the second-order character of the ferroelectric phase transition. 
The presence of a hysteresis loop in the natural optical activity measurements can be associated with the \textit{gyrotropic} order associated with the $P3$ space group~\cite{C_Konak_1978, vlokh1987}, which can indeed be associated with optically active polar domains. 
Furthermore, the linear electrogyration~\cite{aizu1964} coefficient of PGO is among the largest ever recorded, with a value $\gamma_{33}$ =(3.1 $\pm$ 0.3) $\times$ 10$^{-11}$ m/V near 450 K in chromium doped conditions (0.8 $\%$)~\cite{zheludev1976,vlokh1987,konak1978,uesu1991}.
Another intriguing aspect of PGO involves its ferroelectric domain walls (DWs). 
Notably, the material has been observed to exhibit the phenomenon of topological bifurcation, as discussed in previous research~\cite{BAK_PGO,PGO_topology}. Additionally, recent experimental findings have identified the presence of antiferroelectric DWs~\cite{conroy2023observation}. These unconventional behaviors in PGO's domain walls present a compelling case for further investigation, mainly through theoretical explanations grounded in first-principles calculations.

While the past five decades have seen considerable experimental work on PGO, theoretical investigations have been mainly confined to mean-field modeling to fit experimental data, only three  
recent exceptions employed density functional theory (DFT)~\cite{VIENNOIS2018461,PhysRevB.108.L201112,conroy2023observation}. 
In this study, we investigate the microscopic mechanisms driving PGO's phase transitions with the help of \textit{ab initio} calculations. 
The paper is organized as follows.
After reporting our technical details for the calculations and method of analysis and the structural information of PGO, we first analyze the paraelectric $P\bar{6}$ phase through density functional perturbation theory (DFPT).
The resulting calculations of phonon dispersion curves and Born effective charges helps us to identify a single unstable polar phonon branch where both the transverse and longitudinal optical (respectively TO and LO) cases are unstable, emphasizing the hyperferroelectric~\cite{garrity2014} character of the polar phase and the soft mode origin of the phase transition.
In the next section, we scrutinize and characterize the ferroelectric $P3$ phase with DFT and DFPT.
The use of symmetry adapted mode (SAM) decomposition of the distortions present in the $P3$ phase helps us to identify the relevant modes at play in the phase transition.
Then, we build a phenomenological model to describe the energy landscape involving these modes that helps us to understand the role of the spin orbit coupling (SOC) to reproduce the second order kind of the phase transition.
Finally, we extend our analysis to quantify the chirality of the 
$P3$ phase and the associated phonon modes by employing the continuous symmetry measures technique~\cite{Zabrodsky1995}. 
This rigorous approach enables us to pinpoint the origin of chirality within PGO and elucidates the underlying reasons for its gyroelectric properties. 
We argue that the chirality of this material is not associated with a phonon angular momentum, contrarily to what observed in recent works~\cite{Ishito2023,Ueda2023}. The further exploration of the AM shows values close to 1 Ha for some bands in the FE phase and a trend of the paraelectric and ferroelectrics angular momentum distributions with respect to the phonon wave vector.
Our findings not only shed light on the intrinsic chiral nature of the 
$P3$ phase but also provide a deeper understanding of the gyroelectric behavior exhibited by PGO, thereby contributing to the broader comprehension of symmetry-breaking phenomena in ferroelectric materials.

\section{Technical Details}

Structural relaxations, energy, and response function calculations have been performed with the density functional theory (DFT) code ABINIT v9.6.2~\cite{gonze2016,gonze2020} and through norm-conserving pseudopotentials from the PseudoDojo project~\cite{pseudodojo} (v0.4).
The generalized gradient approximation (GGA) with the Perdew-Burke-Ernzerhof functional for solids (PBEsol) flavour~\cite{Perdew1996} have been used. 
A 3x3x3 k-point grid and a cutoff of 50 Ha (1360.57 eV) were employed and found sufficient to converge total energies, structural relaxation (cell parameter and atomic positions), as well as phonon frequencies. 
Density functional perturbation theory (DFPT)~\cite{gonze1997}
response functions - in both the PE and FE phases - were used to obtain the phonon frequencies, the Born effective charges and the permittivity tensor.~\cite{gonze1997}
The Born effective charges and the dielectric tensor allows us to evaluate the non-analytical (NA) dipole-dipole long-range part (LR) of the dynamical matrix~\cite{gonze1997} giving rise to the longitudinal optical (LO) modes.
The phonon dispersion curves were calculated by interpolating the interatomic force constants (IFCs) from the unit cell only by splitting the LR part from the rest (considered the short-range SR).
This interpolation from the unit cell allows us to reduce the computational workload, and it is an acceptable approximation as the unit cell is already large enough (around 10$\times$10$\times$10 \AA$^3$) 
to give reasonable values for the SR part over several neighbors.
The Berry phase theory~\cite{king-smith1993} was employed to obtain the polarization of the ferroelectric phase. 
The latter is also estimated from the Born effective charges to check whether it is free from spurious quanta~\cite{spaldin2012}.
SOC has been included in all the calculations as it appeared to have a surprisingly strong effect in PGO (see ~\cite{supp} and ~\cite{PhysRevB.108.L201112}).
Finally, the group theory analysis of the structural distortions was performed by means of the AMPLIMODES software~\cite{amplimode}.
Because of the size of the system (57 atoms), we will forward the reader to look at the supplemental materials for some extra details and data (like the Born effective charges, the full phonon dispersion and frequencies, etc) that would weight the main text down.

\section{Structural information}

\begin{figure}[htb!]
 \centering
    \includegraphics[width=.9\textwidth]{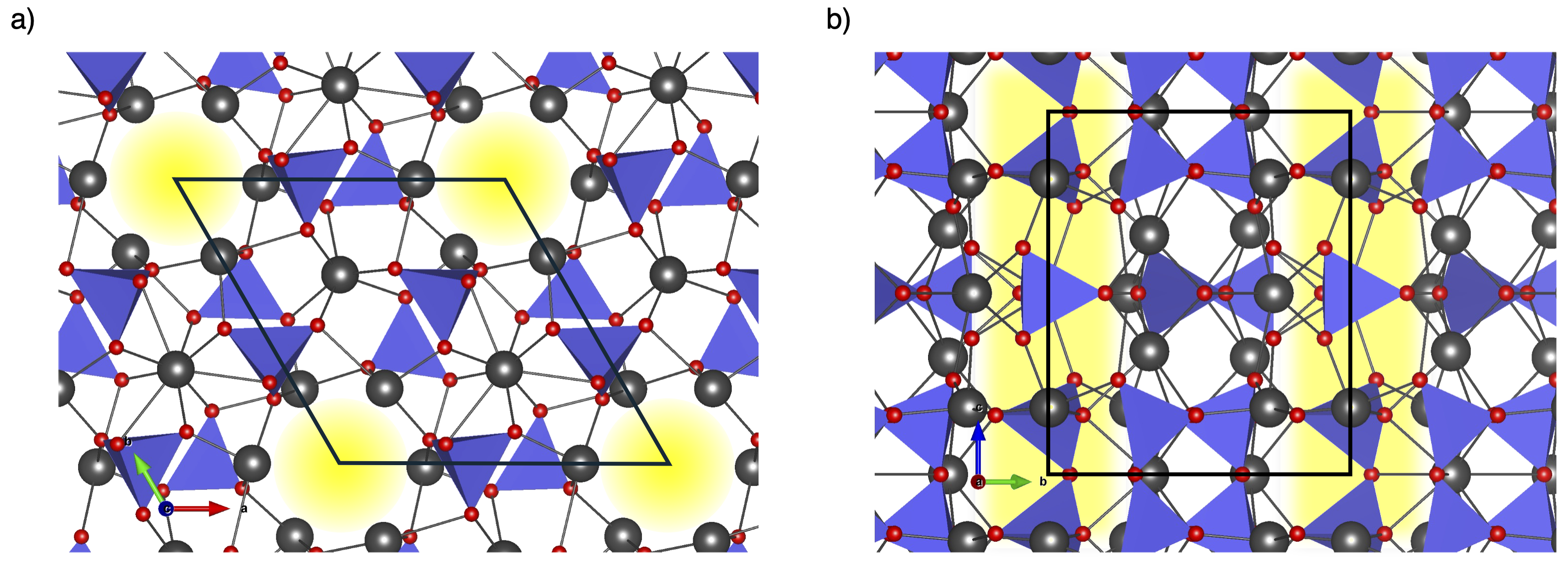}
    \caption{(a) [001] top view and (b) [100] side view of the PE unit cell of Pb$_5$Ge$_3$O$_{11}$. Pb and O atoms are shown in dark grey and red, respectively. GeO$_4$ polyhedral environments are represented in purple. Empty channels are evidenced in yellow.}
 \label{fig:PE}
\end{figure}

Lead germanate is a large band-gap insulator that undergoes a structural phase transition at 450 K. 
It features a high-temperature hexagonal PE $P\Bar{6}$ phase. 
In contrast, its low-symmetry FE $P3$ phase disrupts the six-fold roto-inversion and the mirror symmetry along the c-axis. 
We detail the relaxed structural attributes of the PE and FE phases in the supplementary materials. 
The unit cell for both phases comprises 57 atoms, with the PE and FE phases characterized by 15 and 23 asymmetric Wyckoff positions (WPs), respectively. 
A schematic representation of the high-symmetry PE phase is provided in Fig.\ref{fig:PE}. 
In this structure, lead atoms occupy two distinct positions: those at 6l and 3k WPs form hexagonal configurations around a vacuum volume when viewed along the [001] direction.
In contrast, the other lead atoms are aligned along the c-axis within the bulk part of the material. 
Germanium atoms, in coordination with surrounding oxygens, form either GeO$_{4}$ tetrahedra or Ge$_{2}$O$_{7}$ bitetrahedra. 
The lead atoms serve as bridges between these germanium-oxygen units.
A comparative analysis of the lattice parameters with respect to the employed DFT functional has been presented in a previous work (see Supplementary Materials of ref.~\cite{PhysRevB.108.L201112}).
It is worth noting that the PBEsol functional plus SOC used in our study slightly overestimates the $a$ and $c$ lattice parameters by up to 0.38 \% and 0.25 \% 
compared to experimental values. 
Hence throughout the rest of the manuscript we use the PBEsol plus SOC calculations as a reference, since they provide the most accurate results compared to other functionals such as PBE~\cite{Perdew1996} and LDA.
In the FE phase, the 3d Wyckoff position is occupied by all the oxygen and germanium atoms. 
In contrast, lead atoms are found in the 3d, 1c, and 1b positions. Interestingly, both $P\Bar{6}$ and $P3$ space groups exhibit axial order and a non-zero piezoelectric tensor~\cite{PhysRevLett.116.177602,Hayashida2020}, less commonly observed in the PE phase of ferroelectric materials.

\section{Analysis of the paraelectric phase}

In this section, we analyze the $P\Bar{6}$ PE phase of PGO to identify and characterize the phonon instabilities.
After the complete structural relaxation, we calculated the phonon frequencies, the Born effective charges, and the permittivity at the $\Gamma$ point through DFPT with and without including the spin-orbit coupling (SOC). 
The irreducible representation at $\Gamma$ is $30A'\oplus27A''\oplus64E'\oplus50E''$, where the $A''$ ($\Gamma_2$) and $E'$ ($\Gamma_3/\Gamma_5$) characters are IR active with polarization along the $z$ and $xy$ directions respectively and the $A'$ ($\Gamma_1$) and $E''$ ($\Gamma_4/\Gamma_6$) are Raman active only. 
From our calculated phonons
we find at $\Gamma$ a single unstable TO phonon mode with a frequency $\omega_{0}$ = 34i cm$^{-1}$ and of $\Gamma_{2}$ representation
that is an infra-red (IR) active mode polarised along the c-axis. 
If we look at the difference between the TO and LO frequencies associated with this unstable mode, we realize that they are very close: 34i cm$^{-1}$ for the TO mode vs 28i cm$^{-1}$ for the LO mode. 
We also find that, contrary to several other structural and electronic properties (see Ref.~\cite{PhysRevB.108.L201112} and the next sections of the present manuscript), the inclusion of SOC in the calculation does not influence much this unstable phonon mode frequency: $\omega_{TO}$ = 37i cm$^{-1}$ and  $\omega_{TO}$ = 31i cm$^{-1}$.
Moreover, 
splitting the square frequencies at $\Gamma$ in a short and long-range contributions~\cite{PhysRevB.55.10355,ghosez1996, bousquet2006} 
reveals the short-range nature of the polar instability with $\omega_{0;SR}^{2}$ = -2235.2 cm$^{-1}$ and $\omega_{0;LR}^{2}$ = 1022.5 cm$^{-1}$.

\begin{figure}
     \includegraphics[width=0.6\textwidth]{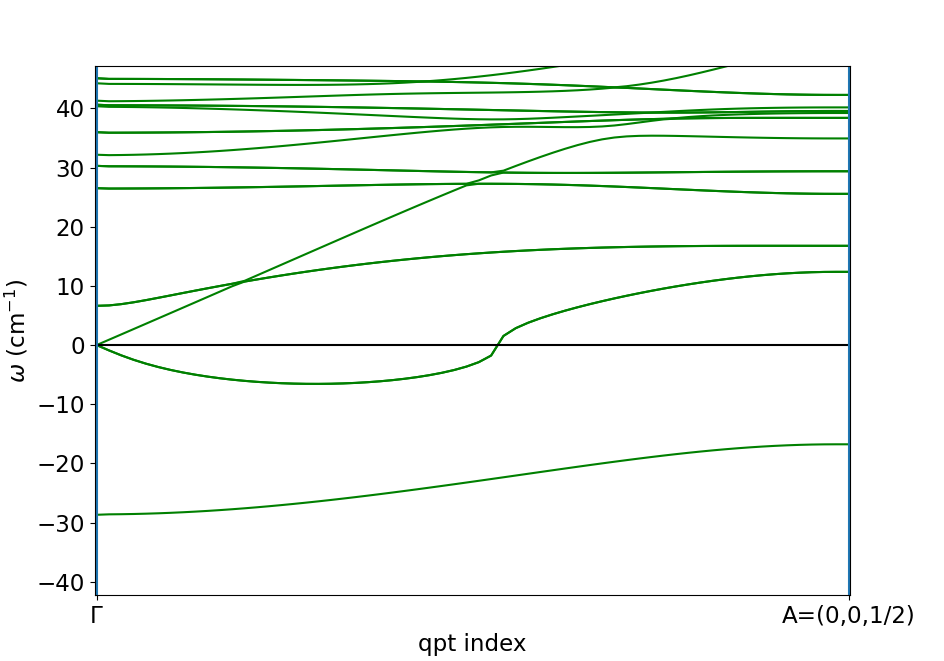}
     \caption{
     Calculated phonon dispersion curves of the PE reference structure of PGO between the zone center $\Gamma$ point and the zone boundary $A$ (0, 0, $\frac{1}{2}$) points and for a zoomed frequency range between $45i$ to $45$ cm$^{-1}$. The NAC correction has been taken into account along the (0,0,1). Hence, the unstable branch running from $\Gamma$ to $A$ corresponds to the LO polar mode at $\Gamma$ associated to the unstable TO mode with $\Gamma_2$ symmetry; The unstable branch from one of the acoustic mode is an interpolation artefact as the condensation of the related elastic instability never gives a lower energy phase. }
     \label{fig:PGO_phonons_PE}
\end{figure}

Given that the unstable mode is polarized along the $c$-axis,
we aim to investigate the complete optical branch extending from the unstable $\Gamma$  point to the $A$ point, which has coordinates (0, 0, $\frac{1}{2}$). 
To facilitate this, we leverage the large size of the unit cell - approximately 10 \AA$^3$ box - to interpolate the phonon dispersion solely based on dipole-dipole interactions at finite $q$-values. 
The outcome of this interpolation between the $\Gamma$ and A points is depicted in Fig.\ref{fig:PGO_phonons_PE}, where we have zoomed to the low-frequency part of the spectrum for clarity.
The presence of a soft acoustic mode in our observations can be ascribed to the constraints of the simplified interpolation model utilized in our analysis. 
Our computational results uncover a comprehensive unstable phonon branch spanning the 
$\Gamma$  to $A$ points of the Brillouin zone, characterized by minimal dispersion. 
This suggests that the model, while effective in a general sense, may not fully capture the complexities of the phonon interactions within the material, indicating the need for a more nuanced approach to accurately represent the dynamic behavior of the phonons across the entire band. To verify that this is not a spurious effect of the interpolation, we have calculated the phonons from DFPT at the $A$ point and found a tiny deviation with respect to the purely interpolated value (17.1i  cm$^{-1}$ from DFPT vs 16.8i cm$^{-1}$ from dipole-dipole interpolation). 
Thus, it is clearly the LO mode that connects the related unstable dispersion branch along the $\Gamma$-$A$ line.

In agreement with the calculations done in Ref.~\cite{conroy2023observation}, our results concerning the soft mode suggests that PGO behaves like a hyperferroelectric, that is a ferroelectric crystal that can support a polarization in $\mathbf{D}$ = 0 condition of the displacement field~\cite{garrity2014}. 
To further probe this possibility, we have computed the Born effective charges  and the electronic permitivity $\varepsilon^\infty$ (see all the tensors in SM~\cite{supp}). 
Even if the band gap is rather large, giving a rather small $\varepsilon^\infty$ ($\varepsilon^\infty_{zz}=5.1$) and even though a few BEC are anomalous (e.g. the $zz$ BEC component of Pb at WP 1i and 1e of around $+4e$ and at 1c close to $+5e$), the overall mean square value $\sqrt{\sum_{jk}(Z^{\alpha}_{jk})^2/57} \simeq (2.7,2.7,2.8)e$ is rather nominal.
Hence, the Coulomb term driving the LO-TO splitting that schematically evolves as the average BEC over $\varepsilon^\infty$ is small enough to keep the LO mode unstable~\cite{garrity2014}. 
Having an average BEC value close to nominal can also means that PGO does not present an overall anomalously large charge transfer due to covalent bonds~\cite{garrity2014}, unlike common ferroelectrics. 
We further show in the Supplementary Materials file~\cite{supp} that the spin-orbit coupling weakly affects the Born effective charges and $\varepsilon^\infty$, i.e. a few percent increase. 
Analysis on known hyperferroelectric compounds such as  LiNbO$_{3}$ and ABC hexagonal systems~\cite{Li2016_LiBO3,khedidji2021} allows to dig deeper into the mechanism behind this short-range instability. 
In particular, Ref.~\cite{khedidji2021} shows that the short-range destabilisation in hyperferroelectrics ABO$_{3}$ and ABC types is coming from negative on-site interatomic force constants (IFCs). However, in the Supplementary file~\cite{supp},  we show that the on-site IFCs are all positive in PGO, therefore ruling out the Ref.~\cite{khedidji2021} argument in the case of PGO. 
While we do not investigate this finding in details, we can possibly understand it from a rigid-unit perspective as reported in, e.g., silicates~\cite{hammonds1996}. 
In fact one can see that both the high and low symmetry phases are comprised of oxygen-sharing Ge- and Pb-centered quasi-rigid polyhedra. 
Thus, the source of the polar instability does not involve individual self-forces (due to the rigidity of the polyhedra), but is rather connected with the stabilisation of oxygen-mediated nearest-neighbour interactions between these units.
The instability would stay geometric but from rigit-unit motions instead of atomic origin~\cite{garcia-castro2014}.


Our calculations, hence, confirms that PGO is a ferroelectric material with a uniaxial polarization and with a proper order parameter as described by the soft mode theory~\cite{cochran1960} and as observed experimentally~\cite{burns1972,hisano1973,hosea1979,satija1982}. 
Further confirmation of a hyperferroelectric character 
would require finding a nonzero polarisation in open circuit boundary conditions. 
Results supporting such conclusion are given in the next section.

\section{Analysis of the ferroelectric phase}

Given that the soft mode aligns with the symmetry of the (FE) phase, it is reasonable to classify the phase transition of PGO as a proper ferroelectric transition. 
We begin our analysis of the polar state with the SOC off.
Starting with the PE structure as a reference, we induce a symmetry-breaking by displacing ions along directions dictated by the unstable phonon eigenvector and with various amplitudes. 
We observe a double-well energy landscape, with energy minima corresponding to a gain of 9.0 meV/f.u.. 
At these minima, we calculate the spontaneous polarization $\mathbf{P}_{s}$ and find a magnitude of 2.8 $\mu$C/cm$^{2}$ along the $c$ axis. 
This value is notably smaller than the experimentally reported 5.0 $\mu$C/cm$^{2}$~\cite{Iwasaki1972} when extrapolated to 0 K. 
To reconcile this discrepancy, we first conduct a structural relaxation with fixed cell parameters and then recalculate the energy gain and polarization. 
The revised values are 61 meV/f.u. and 5.3 $\mu$C/cm$^{2}$, respectively. Consistently with the phonon instability, we detect no in-plane components of $\mathbf{P}_{s}$.
A second structural optimization - in which the relaxation of the lattice parameters with respect to the paraelectric phase is allowed - finally gives an energy gain of 68 meV/f.u. and $\mathbf{P}_{s}$ of 5.9 $\mu$C/cm$^{2}$ along the $z$ direction which extrapolates correctly the experimental value.
Unlike the previous cases, a small in-plane polarisation $\mathbf{P}^{xy}_{s}$ = (-0.07,0.05) $\mu$C/cm$^{2}$ is now observed in the fully relaxed case,
which is associated with the piezoelectric nature of the high symmetry reference structure.
The Berry phase computed values of the polarisation are consistent with those found from the atomic displacements and the Born effective charges.
As we remind that the spin-orbit has been deactivated during the calculations of the aforementioned quantities, from a previous work~\cite{PhysRevB.108.L201112} we also know that the SOC renormalises the ferroelectric barrier up to $\sim$ 30 \% of its value. 
In fact, the SOC increases the unrelaxed soft mode energy up to 13 meV, the ion-only relaxed energy up to 80 meV and the fully relaxed FE ground state energy up to 89 meV.
It is therefore comes natural to understand how $\mathbf{P}_{s}$ is affected by the spin-orbit interaction as well. 
The calculation of the polarization with the SOC included gives $\mathbf{P}_{s}=$ (-0.04, 0.02, 5.5) $\mu$C/cm$^{2}$.
Given that the energies are much affected by the SOC, having a relatively SOC independent polarisation means that much of the SOC contribution affects the non-harmonic part of the energy, a fact that is also supported by the energy of the unrelaxed soft modes computed either with or without spin-orbital contribution. 

The fully relaxed ferroelectric cell parameters are $a=b= 10.257$ \AA\ and $c=10.689$ \AA\ 
align well with experimental findings, as detailed in the supplementary materials~\cite{supp}. 
To check whether a further symmetry lowering might occur, we recalculated the phonon frequencies at the $\Gamma$ point and found no unstable mode, confirming the stability and ground state of our relaxed $P3$ phase.

Hence, we can summarise our DFT calculations as follows. Internal ion relaxation is clearly of paramount importance in reaching both the minimum energy configuration and reproducing the experimental polarisation, while strain optimization has a secondary importance (the additional gain of energy when including the strain is much smaller than the gain of energy given by the internal atom relaxation alone). 
The role of the internal forces can be somehow expected, given the presence of both a large number of atoms (57) in the unit cell with a low symmetry site, and of Pb$^{2+}$ cations that have 6s-6p lone pairs, which are known to lead to a strong relaxation effect in ferroelectric perovskite compounds like PbTiO$_{3}$~\cite{https://doi.org/10.1002/adts.201900029}.

To further understand the PE-FE ground state distortions $\ket{\delta}$ we perform a symmetry adapted mode (SAM) analysis. 
We find that the distortion is characterized by two SAMs. 
The main one (1.47 \AA) is associated with a $P3$ isotropy subgroup and a $\Gamma_2$ irreducible representation (IR) (namely the same as the unstable polar mode). 
The second SAM (0.35 \AA) is instead a mode with $P\bar{6}$ space group and $\Gamma_{1}$ representation. 
This confirms that it is the $\Gamma_{2}$ SAM that breaks the PE symmetry and favors a polar state, while the presence of $\Gamma_{1}$ deformations means that the degrees of freedom already present in the $P\Bar{6}$ phase (i.e. the atoms with WP that are not at high symmetry positions) change to accommodate the polar deformation.

\begin{figure}
    \includegraphics[width=.9\textwidth]{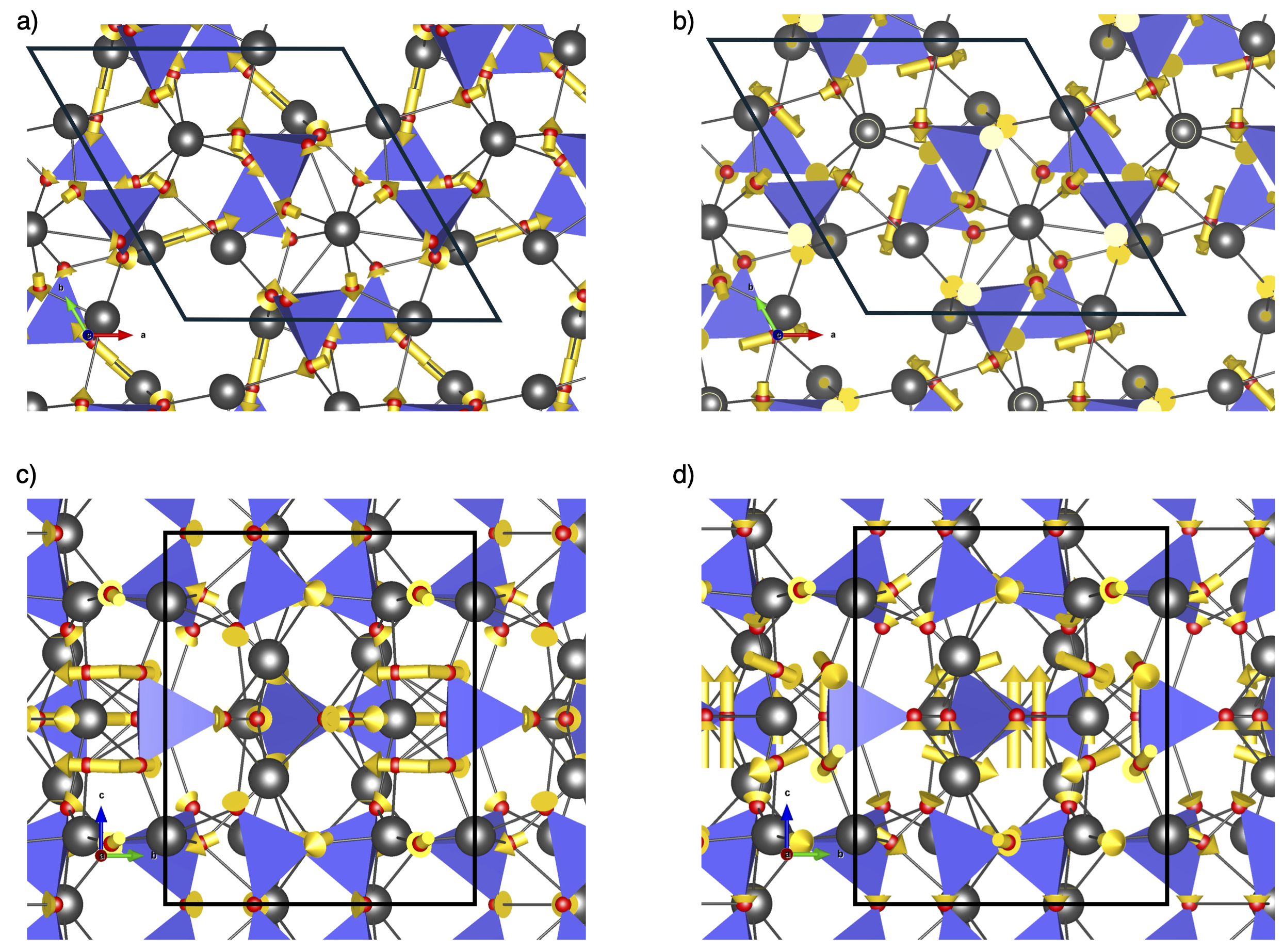}
    \caption{(a)-(c) top and side views for $\Gamma_{1}$ symmetry adapted modes (SAMs), and (b)-(d) for $\Gamma_{2}$ SAMs as calculated with AMPLIMODE software from our relaxed $P\bar{6}$ and $P3$ phases. The arrow lengths are proportionals to the magnitude of the distortions.}
\label{fig:SAMs_VESTA}
\end{figure}

A graphic representation of the two relevant SAMs is given in Fig.~\ref{fig:SAMs_VESTA}. In particular, the invariant mode is clearly constituted by atomic in-plane distortions, while the polar mode contains out-of-plane displacements as well.
To have more details about which phonon modes contribute to the total ground state distortion, we did a projection of the ferroelectric distortion $\bra{\delta}$ into the phonon eigenvectors $\ket{\xi_{i}}$ obtained in the PE phase.
This projection reveals that despite the strong relaxation, the overlap coefficient $\bra{\delta}M\ket{\xi_\text{soft}}$ of the soft mode is about 0.90.
As the normalization is $\bra{\xi_{i}}M\ket{\xi_{j}}$ = $\delta_{ij}$, it means that all the other modes give a total overlap of about 0.44.
Hence, the final ferroelectric distortion is close to the unstable polar mode eigendisplacement but other higher frequency mode eigendisplacements contribute too.
If we now decompose $\ket{\delta}$ into atomic type, we find that O and Pb atoms are those that contribute the most to the polarization.
This result strengthens a rigid unit picture of the phase transition, with $\ket{\delta}$ described by the motion of Ge-O tetrahedra~\cite{Cheng_2018}.

\begin{figure}
     \includegraphics[width=0.8\textwidth]{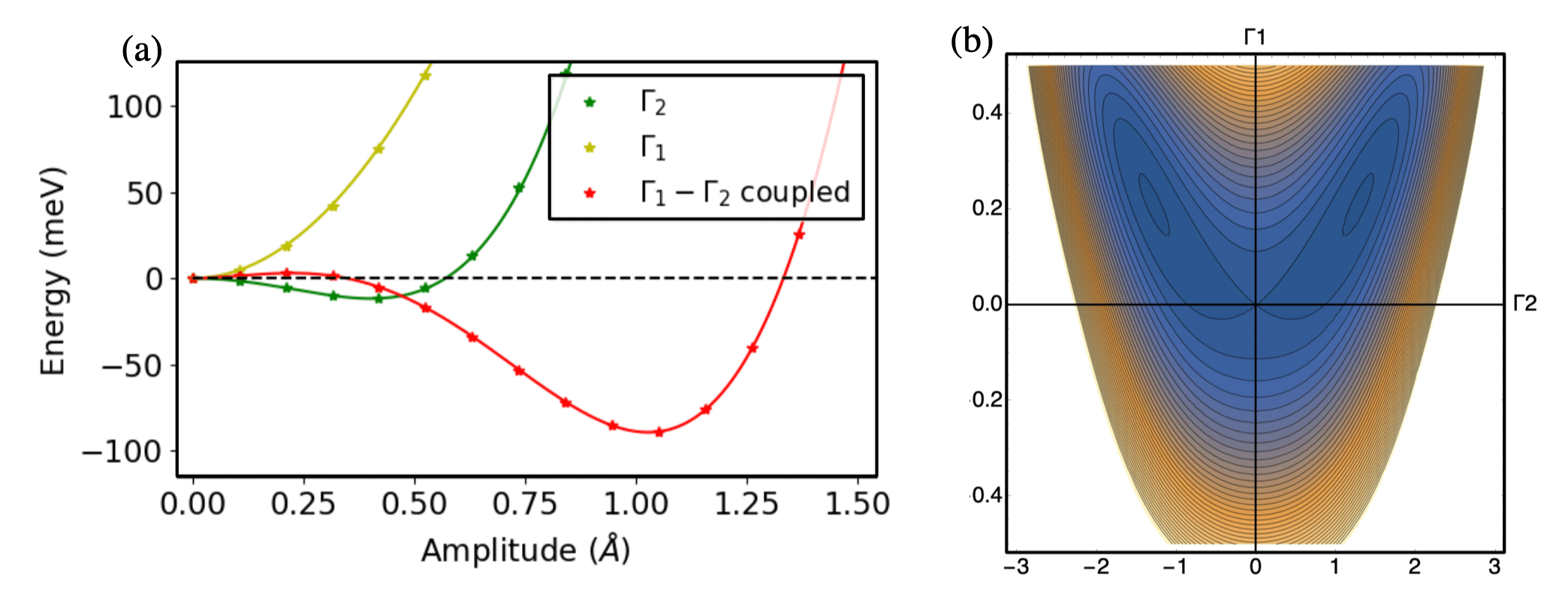}
     \caption{Energy landscape of PGO as a function of the amplitude of the SAM $\Gamma_1$ and $\Gamma_2$. Symbols are DFT calculations while the plain lines correspond to fitted model Eq.\ref{Eq:Umodel}. SOC is included the PE and FE structures have been fully relaxed.}
     \label{fig:SOC_landscape}
     \end{figure}

\begin{figure}
  \includegraphics[width=.6\textwidth]{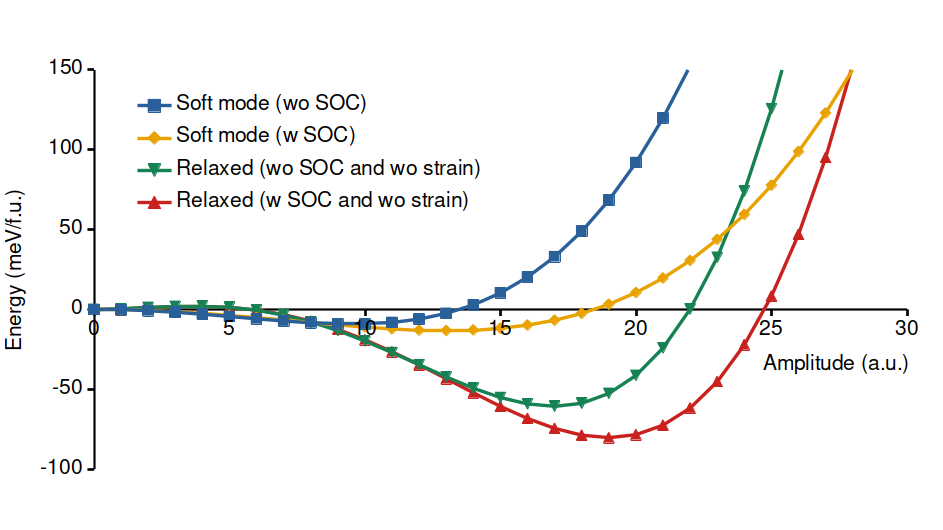}
    \caption{Energy change versus amplitude of the polar unstable mode eigendisplacements and of the  relaxed ferroelectric distortion (both at fixed cell parameter of the PE phase) with and without SOC. The zero energy reference corresponds to the PE $P\bar{6}$ phase}
\label{fig:landscape_w_wo_SOC}
\end{figure}

With the goal of probing the energy landscape (taking again the PE structure as reference), we extract configurations corresponding to either the total $\ket{\delta}$ deformation and its $\Gamma_{1}$ and $\Gamma_{2}$ SAMs projections, linearly interpolating between the high and low symmetry structures. 
Then, we performed DFT calculations as a function of the mode amplitudes. 
To understand the numerical results we 
build a simple phenomenological internal energy model with the $\Gamma_1$ and $\Gamma_2$ SAM mode distortions as order parameters of the system. 
Taking into account the symmetry of these two order parameters, 
we find (see \cite{supp} for a thorough analysis of the fit) that the \textit{ab initio} energy-configuration dataset is well represented by the following expression:

\begin{align}\label{Eq:Umodel}
U(Q_1,Q_2) &\ = \frac{\alpha_1}{2}  Q_1^2 + \frac{\alpha_2}{2} Q_2^2+\frac{\beta_2}{4} Q_2^4  + aQ_{1}Q_{2}^{2} + V_{\text{high}}(Q_1,Q_2)
\end{align}  

where Q$_{1}$ and Q$_{2}$ are the amplitudes of the $\Gamma_1$ and $\Gamma_2$ distortions respectively and where $V_{\text{high}}$ contains higher order terms of the expansion (see Supplemental Material~\cite{supp}).
Since $\Gamma_1$ is invariant under all the symmetry operations of the PE reference, it can appear in all orders (except in the first order, since the forces in the PE reference are zero). 
After fitting the model onto our DFT calculations, we can observe the energy wells and landscape in Fig.~\ref{fig:SOC_landscape}.
We can see that, as expected, the $\Gamma_1$ SAM alone gives a single well around the PE reference (yellow curve) and the $\Gamma_2$ mode alone gives a double well shape (green curve).
When both $\Gamma_1$ and $\Gamma_2$ are coupled together we can observe a strong increase of the double well energy and distortions amplitude (red curve). 
This confirms that, in PGO, the number of internal degrees of freedom strongly enhances the development of the polar distortion into the structure, as anticipated by our previous relaxation procedure. 
This enhancement is mainly due to the attractive $aQ_1Q_2^2$ coupling term, which strongly renormalizes and reduces the value of the anharmonic parameter $\beta_{2}$.  
Nevertheless, our calculations also show that the reduction is not strong enough to affect the sign of the quartic $Q_{2}^{4}$ coefficient, which means that the phase transition remains of the second order kind, as experimentally observed. 
The inclusion of higher order terms ($Q_{1}^{3}$, $Q_{2}^{6}$, $Q_{1}^{2}Q_{2}^{2}$, $Q_{1}Q_{2}^{4}$, etc.) can improve the fitting for bigger values of the SAMs amplitudes, but this has a quite marginal importance near the energy minimum. 
The strong renormalisation of the ferroelectric barrier induced by the spin-orbit interaction has been anticipated by us in a previous work~\cite{PhysRevB.108.L201112}. 
To highlight this effect, we show in Fig.~\ref{fig:landscape_w_wo_SOC} the difference in the energy wells when condensing the soft mode and the different amplitude of the fully relaxed distortion with and without SOC.
We can clearly see the relevance of the SOC on the polar distortion energy landscape where it is mostly within the large atomic displacements that the SOC is at play, and, hence, into the anharmonic part as abserved from the phonon frequencies calculations from DFPT, which were slightly affected but the SOC.

Although we did not probe the Berry phase as a function of the symmetry adapted mode amplitudes, we can safely assume a linear relation $Q_2\sim\mathbf{P}$ which is generally justified for small distortion amplitudes. 
Finally, we stress that although the Berry phase was calculated in $\mathbf{E}$ = 0 conditions, the detection of an unstable LO branch prompts the question about what $\mathbf{P}(\mathbf{D}=0)$ may be. 
This calculation from first principles is prohibitively costly in the state of the art of DFT codes implementation, however, if we assume that the electrostatics affects only the quadratic part of the electric enthalpy, we can estimate $\mathbf{P}(\mathbf{D}=0) \approx \mathbf{P}(\mathbf{E}=0)\frac{\omega_\text{LO}}{\omega_\text{TO}}$ $\sim$ 4.0 $\mu$C$\cdot$cm$^{-2}$, which is a remarkably small reduction in comparison with hyperferroelectric materials such as LiBO$_{3}$ (with B = V, Nb, Ta and Os)~\cite{Li2016_LiBO3}. 
Naturally, a further decrease may be expected if non-harmonic effects are taken into account. 
Given that the previous linear relation actually fails even in the simpler cases 
of ABC hyperferroelectrics~\cite{garrity2014}, we can consider the given number as an upper bound.
Equivalently, the $\mathbf{D}=0$ correction to the free energy brings an additional positive term proportional to P$^{2}$, and it can be observed that the ferroelectric instability is not suppressed as a result of the large dielectric constant of PGO.
Indeed, the most prominent contribution from the electrostatic energy associated with the $\mathbf{D}$ = 0 condition adds a P$^2$/2$\epsilon$ term to the free energy expansion (see Supplementary Materials~\cite{supp} and refs.~\cite{PhysRevB.107.094108,PhysRevB.99.104101} therein included), and this increases the $\alpha_2$ coefficient of Eq.~\ref{Eq:Umodel} from the value of -135.0 meV/$\AA$ to that of -119.5 meV/$\AA$, which is compatible with hyperferroelectricity as suggested in the previous section and in agreement with the recent results reported in ref.~\cite{conroy2023observation}.
Our calculations also show that the effect of the SOC on the energy landscape mainly affects the anharmonic terms via ion relaxation, despite not changing the qualitative picture.

It may come to attention that recent experimental and theoretical results (obtained through piezo-response force microscopy (PFM) and phase-field modeling) and highlighted in Refs.~\cite{BAK_PGO,PGO_topology} have shown the presence of charged head-to-head/tail-to-tail (HH/TT)  domain walls in PGO. 
It has been argued that charged DWs should not form as they would be difficult to screen: this point of view has been justified on the ground that the electronic band-gap of $\sim$ 3 eV is too wide to support a total screening of the depolarization field. 
Moreover - while the presence of some $n$- or $p$- type doping can be expected from, e.g., vacancies in the system - both the VBM and CBM states of PGO have been found to be localized in a recent theoretical work~\cite{PhysRevB.108.L201112}. 
This electronic localization should make the screening of P$_{z}$ at the domain wall by free charges coming from dopants even more difficult on top of the large band gap.
It is thus clear that the screening originates by a different mechanism.

Following this idea, more recent PFM measurements have been explained in terms of a complex topological pattern consisting of bifurcated domains, so that the interface bound charge is practically zero - namely $\rho = \rho_{z} + \rho_{xy} \sim $ 0 with $\rho_{z} = -\partial{P}_z/\partial{z}$  
and $\rho_{xy} = -\partial{P}_x/\partial{x} - \partial{P}_y/\partial{y}$ - which would make $\mathbf{P}$ divergenceless.
In other words, if the variation of the polarisation along $z$ generates a $\rho_z$ charge density, this would be readily compensated by the in-plane variations of P$_x$ and P$_y$.
However, a non-zero hyperferroelectric polarisation in open circuit boundary conditions may be associated with a gap-closing - as also obtained for the LiBeSb and LiNbO$_{3}$ compounds~\cite{Liu_2017} 
and well-described by the simple relation $E_{\text{gap}}(L) = E_{\text{gap}}(0) - 2eLP(D=0)/\epsilon$ ($\epsilon$ being the dielectric constant in the material, $L$ 
the domain size and $E_{\text{gap}}(0)$ the bulk gap) - thus providing for a complete screening of the depolarization field.
Clearly, this is not the observed mechanism in PGO, nevertheless populating the conduction bands may still have a strong impact on the DWs physics. 
For one thing, the ferroelectric barrier is enhanced under $n$-doping conditions~\cite{PhysRevB.108.L201112}. Secondly, the population of the localized CBM cavity states~\cite{PhysRevB.108.L201112} via photo-excitation or electron injection may produce an additional increase of $\partial{P}_{z}/\partial{z}$ assuming the $\mathbf{\nabla}\cdot\mathbf{P} = 0$ relation to be topologically protected. 
This condition could potentially correspond to the realization of the antiferroelectric DWs observed under electron beams~\cite{conroy2023observation}, despite the latter being energetically unfavorable with respect to a sharp domain wall.

\section{Chirality measure and phonon angular momentum}

The phenomenon of gyrotropic switching in PGO has been attributed to the presence of both Ge$_{2}$O$_{7}$ and GeO$_{4}$ units within the same unit cell, a rare occurrence in crystals. 
Neutron diffraction~\cite{iwata1977} and high-resolution transmission electron microscopy~\cite{Jun_Hatano_1997} experiments support the idea that the polarisation arises as a consequence of the twist of Ge$_{2}$O$_{7}$ quasi-rigid units and polar motion of Pb$^{2+}$ cations. 
As the latter form a bridge between the bi-tetrahedra and the GeO$_{4}$ units - via Pb-O bonds - the polar instability generates a rotation of the germanate tetrahedra, which in turn plays a fundamental role in determining the structural chirality. 
The $\Gamma_{2}$ mode can be appreciated in Fig.~\ref{fig:SAMs_VESTA}, and one can indeed see that it produces a chiral twist of the central GeO$_{4}$ elements and remove the mirror plane at z = 0.5.

A compelling theoretical inquiry centers on how individual phonon modes within the infrared (IR) spectrum, associated with the $P3$ isotropy group, contribute to the chirality and gyrotropic switching behavior of PGO. 
To offer a clearer understanding of this behavior, an appropriate metric for chirality must first be established. 
We stress that, contrarily to what is stated in ref.~\cite{conroy2023observation}, the $\Gamma_{1}$ mode cannot be chiral since it is an invariant of a space group which contains a mirror operation. 
Rather and as already mentioned we can attribute an axial symmetry to that mode which survives the phase transition.
Given that the low-symmetry phase is polar and chiral at the same time while the high-symmetry reference phase is paraelectric and achiral, the spontaneous polarization may be a fitting metric or order parameter in this case.
Therefore one may eventually use the mode effective charges~\cite{gonze1997,PhysRevLett.72.3618} 
as a means to probe the chirality of each phonon mode. 

Nevertheless, to quantify chirality we adopt a more geometric approach and thus employ the definition of continuous chirality measure (CCM) proposed by Zabrodosky and Avnir~\cite{Zabrodsky1995}, of which we give a short description as follows. 
The starting point is a chiral distribution Q of N atoms $\{q_{i}\}$. 
We define the following quantity:

\begin{equation}
    \chi_{Q}(G) = 100\times\min_{G\in S_{n}}\Bigg[\frac{\sum_{i}|\vec{q}_{i}-\vec{p}_{i}|^{2}|}{\sum_{i}|\vec{q}_{i}-\vec{q}_{0}|^{2}|} \Bigg] ,
\end{equation}

where $\vec{q}_{0}$ is the geometric center of the reference Q-structure and $\vec{p}_{i}$ are the unknown coordinates of a distribution P of N points with symmetry operations given by the achiral point group G (among the improper rotations S$_{n}$) of choice. 
The P-distribution can be obtained by applying the operations of G to the Q-structure, as described in ref.~\cite{Zabrodsky1995}. 
The structural chirality is thus calculated by searching for the closest (distance-wise) non-chiral distribution of points P$_{\text{min}}$ which preserves the connectivity of Q and is compatible with the symmetry operations of G. 
Clearly, $\chi(Q) = 0$ if Q is achiral. 
On the other hand, $\chi(Q) = 100$ can be shown to be the maximum possible distance with respect to the non-chiral reference structure, which corresponds to the - unrealistic - case of all points of P converging to $\vec{q}_{0}$. 
It is thus realized that the CCM is conceived to quantify the geometrical chirality of a molecule without the foreknowledge of an eventual achiral phase (reachable without breaking interatomic bonds). 
Naturally this is not the case of PGO:
if we condense a $\Gamma_{2}$ mode on the paraelectric equilibrium phase while keeping the amplitude of the deformation small, the achiral reference \textit{must} be the original $\text{P}\Bar{6}$ structure, meaning that we can simply write: 

\begin{equation}
    \chi(\Gamma_{2}) = 100\times\frac{\sum_{j}|\delta^{\ket{e_{k}}}_{j}|^{2}}{\sum_{i}(x_{i;\text{FE}}-x_{\text{C.o.M.};\text{FE}})^{2}} \approx 100\times\frac{\sum_{j}|\delta^{\ket{e_{k}}}_{j}|^{2}}{\sum_{i}(x_{i;\text{PE}}-x_{\text{C.o.M.};\text{PE}})^{2}},
\end{equation}

where we assume $\delta^{\ket{e_{k}}}_{j}$ to be the Cartesian deformation of the m-atom induced by the condensation of the $k$-th polar eigendisplacement. 
Clearly, the previous expression can be employed for phonon modes and SAMs as well, provided that they belong to the representation with $P3$ isotropy space group. 
We have thus calculated $\chi$ for PGO in several relaxation conditions, with the results reported in tabs.~\ref{tab:chiraltab}. The invariant distortion that comes with optimising the ferroelectric structure must be subtracted in order to compute the CCM.

\begin{table}[h]
\centering
\begin{tabular}{c|c|c|c|c}
    & PE  & FE(unstable) & FE (I) & FE (I+V) \\
    \hline
 $\chi$($\Gamma_{2}$) &  0.0 & 0.03 & 0.04(0.14) & 0.04(0.16)\\
 $\chi(Q)$ &  0.0 & / & 0.04(0.55) & 0.04(0.59) \\
 $\Delta{E}$ (meV)& 0.0 & -13 & -80  & -89 \\ 
 \end{tabular}
\caption{Chirality measure for the $P\bar{6}$ (PE) and $P3$ structures. 
FE (unstable) corresponds to the freezing of displacements associated to the unstable phonon mode eigendisplacement (i.e. no relaxation is done but the calculation is done at the lowest energy point); I corresponds to the ion relaxation only (i.e. the cell parameters are not relaxed and fixed to those of the $P\bar{6}$ PE reference); I+V corresponds to the fully relaxed case of both cell parameters and ionic positions. 
$\chi$(Q) and $\Delta$E have been computed with the spin-orbit interaction on. 
The numbers between parentheses represent the (normalised) euclidean distance between the two phases when the achiral $\Gamma_1$ contribution is not subtracted.}
\label{tab:chiraltab}
\end{table}

We can see that the CCM of the soft eigenmode is substantially unaffected by the relaxation process, although the euclidean distance between the two phases is about five times bigger.
This is in agreement with the existence of a strong coupling between the soft and invariant modes and, at the same time, of a weak overlap between the distortion and stable eigenstates with $\Gamma_2$ representation. 
While the CCM of the analysed chiral/polar mode is small, there are in total 171 phonon modes many of which are also chiral. 
It is thus natural that we look at the value of $\chi$ for all the phonon modes at the $\Gamma$ point.

 \begin{figure}
     \includegraphics[width=0.6\textwidth]{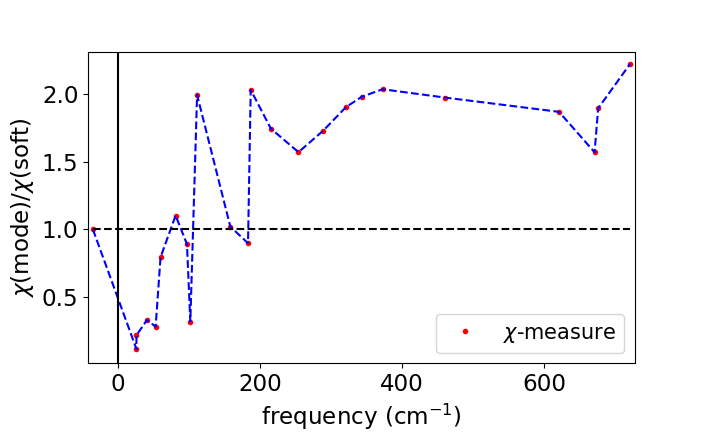}
     \caption{Phonon mode chirality (as calculated from CCM method) as a function of the frequency (the dashed line is a guide to the eye and is at the amplitude of the unstable mode value). 
     The soft mode value is taken as a normalization reference.
     Only the polar modes with $\Gamma_{2}$ character 
     are considered (modes polarized along the $z$ direction). }
     \label{fig:phonon_chirality}
     \end{figure}

The result is highlighted in Fig.~\ref{fig:phonon_chirality} where we show the mode chirality versus phonon frequency as calculated with CCM applied on the eigendisplacement vectors.
We can see that several high-frequency modes have a mode chirality that is larger than the one of the unstable mode driving the phase transition ($\chi > \chi_{\text{unstable $\Gamma_2$}}$) while low-frequency modes have the tendency to have a smaller value. 
Hence, even if the unstable polar mode gives by far the largest overlap projection ($\bra{\delta}M\ket{\xi}_\text{soft}\sim{0.9}$) onto the full distortion, the small extra polar modes coefficients that are at play to completely characterize the polar deformation 
also possess an enhanced CCM.
These strongly chiral modes are mostly associated with  oxygen atom vibrations as their lower mass makes them contributing the most to high frequency vibrations.
Additionally and with respect to eq.\ref{Eq:Umodel}, we can conclude that the equilibrium structural chirality of the ferroelectric phase of PGO is tight to the Q$_{2}$ amplitude and do not depend considerably on the interaction between polar and invariant SAMs, unlike the spontaneous polarisation.

The high-frequency strongly chiral modes only weakly affect the behavior of the phase transition, however, it could be advisable to design non-equilibrium strategies to couple those states to the polar distortion, since it would guarantee a significant level of control on the already remarkable gyrotropic properties of PGO, with the possibility of realizing a ferrochiral memory device.
Recent theoretical works have devised a mechanism based of infrared pumping to obtain fast polarisation reversal in ferro-distorted perovskites~\cite{PhysRevB.92.214303}. 
A subsequent experimental verification~\cite{PhysRevLett.118.197601} on the rhombohedral phase of LiNbO$_{3}$ has found only a partial (without reaching the reversed equilibrium value) and temporary switching, with the effect of the reversal being canceled after a transient. 
The reasons behind the incompleteness (only $\sim$ 40 $\%$) if the switching has been attributed to spatial inhomogeneities, while its cancellation - with a return to the original state after some time after the initial pump - has been explained in terms of coupling with other modes, not considered in ref.~\cite{PhysRevB.92.214303}, and in terms of missed relaxation along the unstable phonons orthogonal to the $c$ axis, given the cubic nature of the high-temperature phase of the material under consideration. 
Chen et al.~\cite{Chen2022} have found that it is possible to achieve a full reversal in a rhombohedral phase co-occurring with an in-plane rotation, but that seems to require a fine-tuning of the amplitude of the pulse.
Thus, they have proposed the realization of a complete and permanent switching through a squeezing mechanism, with the high and low symmetry phases being tetragonal and orthorhombic respectively. 
Assuming an initial P$_{z} \neq$ 0, a laser pulse along the $z$ direction is used to cancel the out-of-plane polarisation and to create in-plane polar distortions (fig. 2 of ref.~\cite{Chen2022}). 
After that, three pulses (equally separated by a time lag) are applied along the $a$, $b$, and finally $c$ crystal axes. 
The final outcome is the full and permanent reversal of P$_{x}$ and P$_{y}$, while P$_{z}$ remains zero (fig. 3 of ref.~\cite{Chen2022}). 
It thus appears that the xy-rotational component is a fundamental prerequisite to achieving fast polarisation switching in a controllable fashion in ferroelectric perovskites. 
On the other hand, PGO is an uniaxial crystal, and the in-plane rotation of $\mathbf{P}$ is energetically unfavoured. 
It is, hence, possible, that a fast switching mechanism as envisioned in Ref.~\cite{PhysRevB.92.214303} could be more easily realized in this system.

Given the recent surge of interest concerning structural chirality in crystals~\cite{fecher2022}, 
we provide a comparison between the kind of chirality as found in PGO and that for instance, related to zone boundary modes as observed in several 2D and 3D systems~\cite{PhysRevLett.115.115502,Ishito2023,Ueda2023}
with the C$_{3}$ symmetry.
The present literature on the topic thus far has been focused on circularly polarised phonon modes, which can in fact be triggered by photons with the same polarisation~\cite{Ueda2023} and produce an orbital magnetic moment as a result of their circular motion~\cite{PhysRevMaterials.3.064405}.
Since the three-fold symmetry is associated with a pseudo-angular momentum (PAM)~\cite{PhysRevLett.115.115502}, selection rules ensue from its conservation. 
While in two dimensions the valley chirality coincides with a local circular rotation (defined per sublattice), in 3D such rotation is combined with a propagation along an axis perpendicular to the rotational plane. 
We further point out that 3D chiral phonons have been defined and observed in enantiomorphic crystals~\cite{Ishito2023}, where the direction of the circulation defines the space group and therefore a handedness~\cite{fecher2022}.
Instead, at $\Gamma$ the modes are static and the system has the symmetry of the point group. 
This means that the circular polarisation of the phonons averages to zero for each mode and that a handedness cannot be defined as in the previously mentioned cases. 
Also, no angular momentum (AM)~\cite{PhysRevLett.112.085503} should be expected at zone center as a consequence of the time-reversal symmetry alone, which is confirmed for each phonon branch by our numerical calculations~\cite{supp}.
Due to the high number of bands we find more practical to analyse the distribution of the AM - the angular momentum density of states - along the in-plane $\Gamma\rightarrow$K (1/3,1/3,0) and out-of-plane $\Gamma\rightarrow$A (0,0,1/2) Brillouin zone directions as shown in fig.~\ref{fig:lz_histo} (we report the full band-by-band computations in the supplementary file~\cite{supp}). 
In particular, it is clear how higher values of the AM are reached in the ferroelectric phase and in particular along the $c$ direction. 
Moreover, values close to 1 Ha are also more frequently reached in the out-of-plane case of the ferroelectric phase.
A closer inspection of the band decomposition of the angular momentum~\cite{supp} also highlights the near zone boundary (centre) character of the in-plane (out-of-plane) FE distribution, with the PE bands behaving in a complementary way.

 \begin{figure}
     \includegraphics[width=0.6\textwidth]{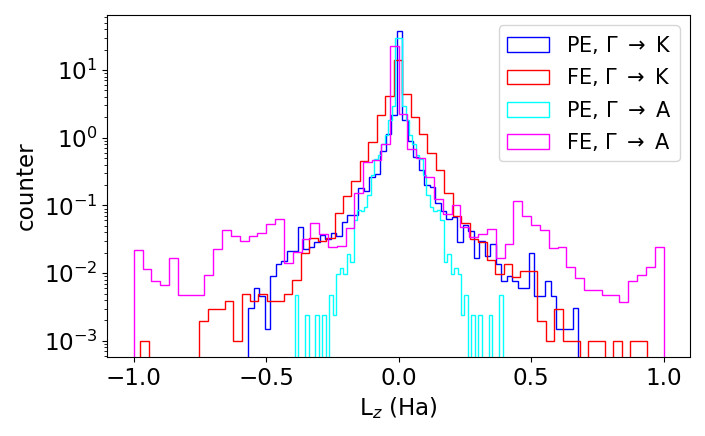}
     \caption{Distribution (log-scale) of the phonon angular momentum along the $\Gamma\rightarrow$K and $\Gamma\rightarrow$A directions.}
     \label{fig:lz_histo}
     \end{figure} 

Therefore, and to conclude this section, the chirality of PGO herein reported has no AM and is thus a property of the point group itself (which polar representation is axial and thus chiral as well) and is not associated with a specific handedness. 
As such it can be triggered by a linearly polarised electric field and its tuning matches that of the spontaneous polarisation.

\section{Summary and conclusions}

We have conducted first-principles calculations to investigate various properties and the phase transition from paraelectric to ferroelectric states in the compound Pb$_5$Ge$_3$O$_{11}$ (PGO). 
Our findings illuminate the microscopic mechanisms underlying the polar deformation, pinpointing a polar mode as the primary source of instability. 
By calculating the Born effective charges, we establish no significant anomalous charge transfer, suggesting that geometric effects or lone pairs are more decisive factors. 
Our data corroborate the classification of PGO as a hyperferroelectric compound~\cite{conroy2023observation}. 
This conclusion is evidenced by the persistence of the instability even under the $\mathbf{D}$ = 0 condition (i.e., after accounting for LO-TO splitting). 
This observation implies that domains polarized along the $c$-axis would remain stable despite variations in boundary conditions, ranging from periodic ($\mathbf{E}$ = 0) to open circuit boundary conditions ($\mathbf{D}$ = 0). 
Furthermore, we demonstrate that the phonon branch associated with the soft mode remains unstable up to the zone boundary.

Exploring the energy landscape reveals the central role of an invariant $\Gamma_{1}$ and a polar $\Gamma_{2}$ modes: their non-linear coupling bolsters the magnitude of the polarisation along the $z$-axis and deepens the energy barrier between opposite domains. 
We hypothesize that this effect - remarkable despite the relatively small magnitude of the invariant distortion - may be associated with the presence of lone pairs of Pb atoms. 
The computed spontaneous polarisation - either with the Berry phase approach or the Born effective charges - is consistent with the value found in experiments. 
Also, we correctly reproduce the second-order character of the phase transition.

We further discuss the behavior of domain walls based on recent results that appeared in the literature~\cite{BAK_PGO,PGO_topology}. 
Having an uni-axial $\mathbf{P}$ (parallel to the $z$-axis) means that domains can meet in a charged head-to-head/tail-to-tail configuration, thus with a sizable depolarisation field: normally, the fulfillment of such requirement would remove or strongly reduce the polar instability~\cite{PhysRevB.73.020103, PhysRevB.83.184104}, if the DW bound charge stemming from $\partial{P}/\partial{z}$ cannot be properly screened. 
However, this does not seem to be the case for PGO since experiments do detect the formation of domains below the critical temperature~\cite{BAK_PGO,PGO_topology,conroy2023observation}.
Therefore, we are left with the question of how charge neutrality can be ensured and the depolarisation field screened. 
A recent theoretical development suggests the formation of an in-plane polarisation, which would neutralize the bound charge associated with $P3$ polar phase at the interface without needing a free carrier (n-p) distribution. 
The resulting $\mathbf{\nabla}\cdot\mathbf{P}$ = 0 condition has a topological character and is associated with domain bifurcation, observed via piezoresponse force microscopy. 
We further conjecture that the topological index associated with the charge neutrality and the recently discovered conduction cavity states~\cite{PhysRevB.108.L201112} in PGO may be used to control the domain walls through the screening effect.

The spontaneous polarisation of PGO also comes with a gyrotropic order, primarily associated with the tilting of the GeO$_{4}$ units concerning the high symmetry configuration. 
We have evaluated the chirality from CCM for the $\Gamma_2$ phonon modes and the relaxed distortions. 
It is found that structural optimization can increase chirality as a result of the interaction between the soft polar mode and the invariant modes. 
Furthermore, the calculation of the CCM associated with $\Gamma_2$ polar phonons shows the presence of high frequency modes with a chirality $\chi$ value twice as high as the value associated with the unstable eigenstate. 
We argue that developing interaction with such modes (e.g. with ultrafast laser excitations~\cite{forst2011}) could produce novel effects in the realm of polarisation switching, with a potential increased level of control on the gyrotropic properties of this material and with the possibility of creating storage devices based on geometric chirality.
It would also be interesting to compute from first-principles~\cite{zabalo2023} the optical activity associated with the phonon modes and the polar distortions to have a clearer idea of the link between chirality (as calculated through CCM) and the optical activity in PGO.
We also highlight that the angular momentum of the polar and chiral $\Gamma_2$ phonons is strictly zero, in contrast with previously reported chiral phonons that are away from zone center with non-zero angular momentum.

Hence, PGO is a versatile materials with numerous properties of high interest for multifunctional applications embedded into a single bulk material, yet not fully explored or exploited.

\section*{Acknowledgements}
The authors acknowledge E. McCabe, Z. Romestan and H. Djani for fruitful discussions, and M. Vaestraete and M. Mignolet for sharing their on-going phonon angular momentum implementation within the ABINIT code (not yet in the production version).
Computational resources have been provided by the Consortium des \'Equipements de Calcul Intensif (C\'ECI), funded by the Fonds de la Recherche Scientifique (F.R.S.-FNRS) under Grant No. 2.5020.11 and the Tier-1 Lucia supercomputer of the Walloon Region, infrastructure funded by the Walloon Region under the grant agreement n°1910247.
MF \& EB acknowledges FNRS for support and the PDR project CHRYSALID No.40003544. Work at West Virginia University was supported by the U.S. Department of Energy (DOE), Office of Science, Basic Energy Sciences (BES) under Award DE‐SC0021375. This work also used Bridges2 and Expanse at Pittsburgh supercomputer center and San Diego supercomputer center through allocation DMR140031 from the Advanced Cyberinfrastructure Coordination Ecosystem: Services \& Support (ACCESS) program, which is supported by National Science Foundation grants \#2138259, \#2138286, \#2138307, \#2137603, and \#2138296.

\bibliographystyle{apsrev4-2}
\bibliography{biblio}

\newpage

\section{Supplementary Material}

\subsection{Phonons, irreducible representation, permittivity and Born effective charges in the PE phase}

\begin{figure}
     \includegraphics[width=0.5\textwidth]{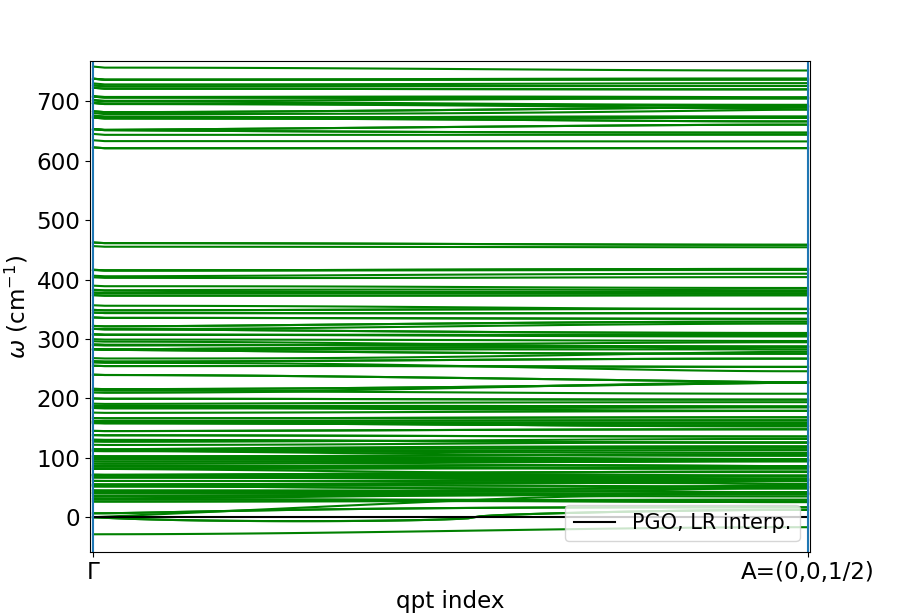}
     \caption{Full phonon spectrum obtained via interpolating a long-range dipole model with DFPT calculations at the $\Gamma$ point.}
     \label{fig:full_phonons}
     \end{figure}

\begin{figure}
     \includegraphics[width=0.5\textwidth]{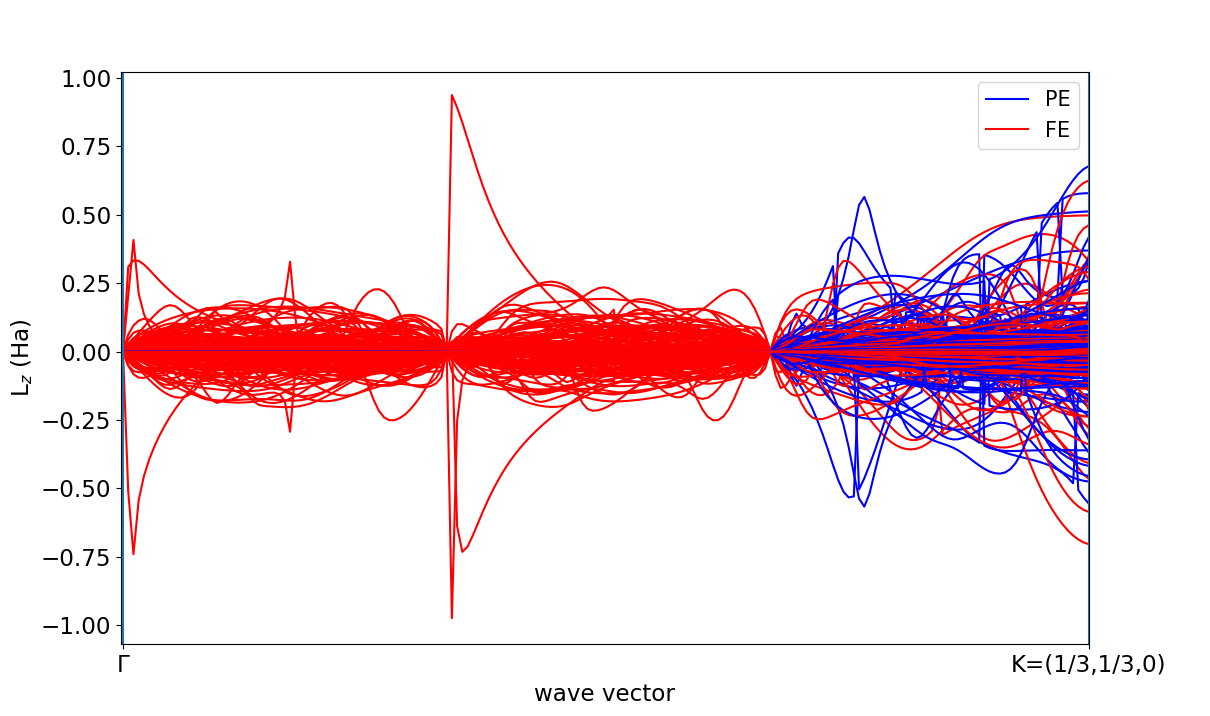}
     \caption{
     Calculated phonon angular momentum between the zone center $\Gamma$ point and the zone boundary $K$ (1/3, 1/3, 0) point.}
     \label{fig:angmom_K}
\end{figure}

\begin{figure}
     \includegraphics[width=0.5\textwidth]{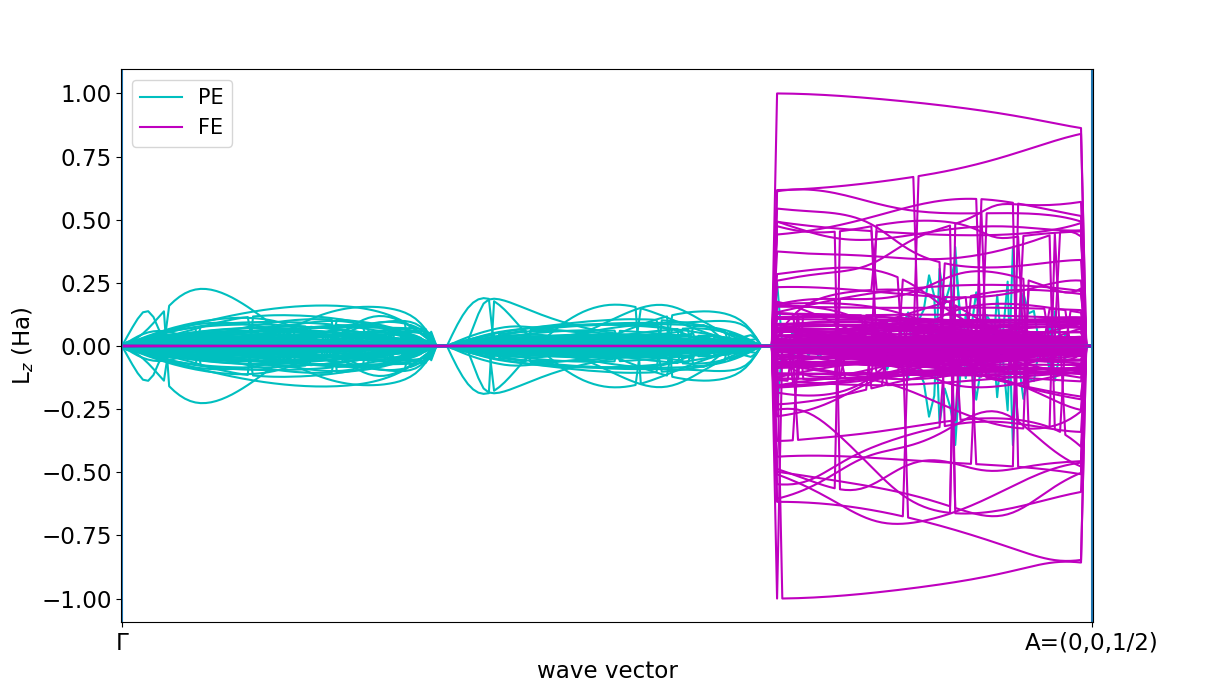}
     \caption{
     Calculated phonon angular momentum between the zone center $\Gamma$ point and the zone boundary A (0,0,1/2) point.}
     \label{fig:angmom_A}
\end{figure}

We report the irreducible representation of the P$\Bar{6}$ phase in tab.~\ref{tab-characP-6} and the complete phonon spectrum in fig.~\ref{fig:full_phonons}. The phonons have been calculated with the dipole model along with DFPT at zone centre. The bands appear nearly flat as a result of the small size of the Brillouin zone of PGO.
While no phonon angular is present at zone centre because of the time-reversal symmetry, it is possible to get a finite AM at finite $\mathbf{q}$.
In figs.~\ref{fig:angmom_K} and~\ref{fig:angmom_A} we report the angular momentum along the $\Gamma\rightarrow$K and $\Gamma\rightarrow$A directions respectively for each phonon mode and for both the P$\Bar{6}$ and P3 phases. 
Considering the in-plane direction, the PE angular momentum tends to be non-negligible only near the zone boundary, while the FE bands possess a non-zero AM much closer to zone centre.
This situation is reversed along the $c$ direction, which shows large FE-momenta closer to the A point and a predominance of the paralectric bands when approaching the $\Gamma$ point.

Among the 171 phonon modes we use the $\Gamma_{2}$ representation to detect the chiral ones and measure their CCM in the main text.
Tab.~\ref{tab:dielec} and~\ref{tab:BEC} instead show the $\epsilon^{\infty}$ tensor and the Born effective charges (BECs). While the dielectric tensor is not much affected by the SOC, it presents a relatively large and anisotropic value along $z$.
Likewise, the Born effective charges are also not much affected by the spin-orbit interaction and they keep close to their nominal values as explained in the main text. 
The large value of $\epsilon^{\infty}_{zz}$ and the small BECs are responsible for the hyperferroelectric character of PGO.

\begin{table}[htbp!]
\begin{center}
\begin{tabular}{lcccccccc}
\hline
\hline
        &                       & $E$ & $3^+$ & $3^-$ & $m$ & $-6^+$ & $-6^-$ & function \rule{0pt}{10pt}\\
\hline
$A'$ & $\Gamma_1$ & 1 & 1 & 1 & 1 & 1 & 1 & $x^2+y^2$, $z^2$, $J_z$ \rule{0pt}{10pt} \\
$A''$& $\Gamma_2$ & 1 & 1 & 1 & -1&-1&-1& $z$ \rule{0pt}{10pt}\\
$E'$ & $\Gamma_3$ & 1 & $w$& $w^2$& 1& $w$ & $w^2$& $(x,y)$ $(x^2-y^2,xy)$ \rule{0pt}{10pt}\\
        & $\Gamma_5$& 1 & $w^2$& $w$& 1& $w^2$&$w$& \\
$E''$& $\Gamma_4$ & 1 & $w$& $w^2$& -1& $-w$ & $-w^2$& $(xy,yz)$ $(J_x,J_y)$ \rule{0pt}{10pt}\\
        & $\Gamma_6$& 1 & $w^2$& $w$& -1& $-w^2$&$-w$& \\
\hline
\end{tabular}
\caption{Character table of the PE space group of PGO.}
\label{tab-characP-6}
\end{center}
\end{table}

\begin{table}[htbp!]
\begin{center}
\begin{tabular}{lcccc}
\hline
\hline
Exc & $\varepsilon^\infty_{xx}$  & $\varepsilon^\infty_{zz}$  & $\varepsilon_{xx}$ & $\varepsilon_{zz}$ \\
\hline
LDA (wo SOC)     & &  &  & \\
PBE (wo SOC)    & &  &  &  \\
PBE (w SOC)     & &  &  &  \\
PBEsol (wo SOC) & 4.7 & 4.9 & 21.0 & NA \\
PBEsol (w SOC)  & 4.9 & 5.1 & 21.9 & NA \\
Exp.            &     &  &  &  \\
\hline
\end{tabular}
\label{tab:dielec}
\caption{Electronic and total permittivity tensor of the PE phase of PGO for different exchange-correlation functionals and with and without SOC.
The total permittivity tensor cannot be calculated along the $z$ direction as an unstable polar mode is present with an imaginary frequency (noted NA).}
\end{center}
\end{table}


\begin{table}[htbp!]
\begin{center}
\begin{tabular}{ccccc}
\hline
\hline
 Atom  & site sym. & wo SOC & w SOC &  \rule{0pt}{8pt}\\
\hline
Pb$_1$ & $3k$ & 
$\begin{pmatrix}
 2.27  &  0.20 &  0.00 \\
 0.36  &  3.95 &  0.00 \\
 0.00  &  0.00 &  3.42 
 \end{pmatrix}
 $
  &  
 $\begin{pmatrix}
  2.30 &  0.24 &  0.00 \\
  0.42 &  4.00 &  0.00 \\
  0.00 &  0.00 &  3.43
 \end{pmatrix}
 $ 
 &
  \\
Pb$_2$ & $6l$ & 
$\begin{pmatrix}
 3.60 &   0.66 &  0.24 \\
 0.74 &   2.38 &  0.13 \\
 0.09 &   0.13 &  2.66
\end{pmatrix}
$
& 
$\begin{pmatrix}
3.63  &  0.65 &   0.24 \\
0.75  &  2.36 &   0.16 \\
0.11  &  0.16 &   2.65
\end{pmatrix}
$
&  \\
Pb$_3$ & $1e$ &  
$
\begin{pmatrix}
 3.09 & 0.19 &  0.00 \\
-0.19 & 3.09 &  0.00 \\
 0.00 & 0.00 &  3.91
\end{pmatrix}
$
& 
$
\begin{pmatrix}
 3.12 &  0.22 &  0.00 \\
-0.22 &  3.12 &  0.00 \\
 0.00 &  0.00 &   4.08
\end{pmatrix}
$
&  \\
Pb$_4$ & $2i$ &  
$
\begin{pmatrix}
 2.68 &  0.04 & 0.00 \\
-0.04 &  2.68 & 0.00 \\
 0.00 &  0.00 & 4.10 
\end{pmatrix}
$
& 
$
\begin{pmatrix}
 2.68 & 0.05 &  0.00 \\
-0.05 & 2.68 &  0.00 \\
 0.00 & 0.00 &  4.17

\end{pmatrix}
$
&  \\
Pb$_5$ & $1c$ &  
$
\begin{pmatrix}
 2.27 & 0.03  & 0.00 \\
-0.03 & 2.27  & 0.00 \\
 0.00 & 0.00  & 4.75
\end{pmatrix}
$
& 
$
\begin{pmatrix}
 2.26  &  0.03  &  0.00 \\
-0.03  &  2.26  &  0.00 \\
 0.00  &  0.00  &  4.95
\end{pmatrix}
$
&  \\
Pb$_6$ & $2h$ & 
$
\begin{pmatrix}
 3.09 & 0.14 & 0.00 \\
-0.14 & 3.09 & 0.00 \\
 0.00 & 0.00 & 3.78
 \end{pmatrix}
$
& 
$
\begin{pmatrix}
 3.12 &   0.15 & 0.00 \\
-0.15 &   3.12 & 0.00 \\
 0.00 &   0.00 & 3.82
 \end{pmatrix}
$
&  \\
Ge$_1$ & $6l$ & 
$
\begin{pmatrix}
 2.88 & 0.28 &  0.01 \\
 0.34 & 4.01 & -0.07 \\
-0.48 & 0.18 &   3.14
 \end{pmatrix}
$
& 
$
\begin{pmatrix}
 2.88 & 0.29 &  0.01 \\
 0.33 & 4.11 & -0.03 \\
-0.51 & 0.21 &  3.16
\end{pmatrix}
$
&  \\ 
Ge$_2$ & $3k$ &
$
\begin{pmatrix}
 3.88 & 0.22 & 0.00 \\
 0.09 & 3.06 & 0.00 \\
 0.00 & 0.00 & 3.09
 \end{pmatrix}
$
&
$
\begin{pmatrix}
 3.92 &  0.24 &  0.00 \\
 0.08 &  3.08 &   0.00 \\
 0.00 &  0.00 &    3.11
  \end{pmatrix}
$
&  \\
\hline
\end{tabular}
\caption{Calculated Born effective charge tensors (unit of the charge of one electron) of nonequivalent Pb and Ge atoms of the PE reference of PGO.}
\label{tab:BEC}
\end{center}
\end{table}





\newpage


\subsection{Force constants model}

\begin{figure}
     \includegraphics[width=0.7\textwidth]{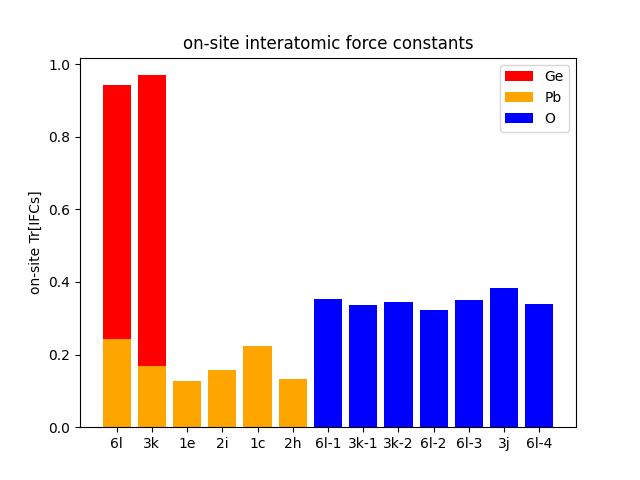}
     \caption{On-site interatomic second order force constants, trace calculated per P$\Bar{6}$ Wyckoff position. Different non-equivalent positions with the same degeneracy and for the same ion (oxygen) are extra-labelled with a number.}
     \label{fig:IFCS}
\end{figure}

The phonon band structure obtained in the main text and in the previous supplementary section has been obtained by interpolating $\Gamma$ calculations via DFPT and a long-range dipole-dipole interaction model. 
In fig.~\ref{fig:IFCS} the trace of the on-site interatomic second order force constants (IFCs) is plotted per Wyckoff position in the paraelectric phase. 
One can clearly observe that all on-site terms are positive definite, a property that is presumably related to the rigid-unit character of PGO. 
Therefore, unlike certain ABO$_{3}$ and ABC hyperferroelectrics~\cite{khedidji2021}, the polar instability is not produced by self-forces but rather by the short-range nearest neighbour interactions.

\subsection{Energies and Landau model}

The amplitudes of the $\Gamma_{1}$ and $\Gamma_{2}$ symmetry adapted modes are shown in tab.~\ref{tab:Ampli_SAMs}, while the relevant ground state energies are reported in tab.~\ref{tab:DeltaE}.

\begin{table}[htbp!]
\begin{center}
\begin{tabular}{lcccc}
\hline
\hline
Exc && $\Gamma_1$  & $\Gamma_2$ &  \\
\hline
LDA (wo SOC)    &&   &   &  \\
PBE (wo SOC)    &&  &   &  \\
PBE (w SOC)     && 0.37 & 1.47 &  \\
PBEsol (wo SOC) && 0.29  & 1.40  &  \\
PBEsol (w SOC)  && 0.32  & 1.47  &  \\
\hline
\end{tabular}
\label{tab:Ampli_SAMs}
\caption{SAMs amplitudes (\AA) as obtained for different functionals and with/without SOC.}
\end{center}
\end{table}

\begin{table}[htbp!]
\begin{center}
\begin{tabular}{lcccc}
\hline
\hline
Exc && $\Delta E$ (no relax)  & $\Delta E$ (w/o strain)  & $\Delta E$ (full) \\
\hline
LDA (wo SOC)    &&  4    & 29  & 34  \\
PBE (wo SOC)    &&  14   & 98  & 105 \\
PBE (w SOC)     &&  21   & 118 & 136 \\
PBEsol (wo SOC) &&  9    & 61  & 68  \\
PBEsol (w SOC)  &&  13   & 80  & 89  \\
\hline
\end{tabular}
\label{tab:DeltaE}
\caption{Gain of energy $\Delta E$  (meV/f.u.) between the PE and the FE phase of PGO for different exchange-correlation functionals and with/without spin-orbit coupling. The "full" label indicates the relaxation of both cell parameters and atomic positions, the w/o strain case means relaxation of the atomic positions with PE fixed cell parameters, and the "no relax" case represents the condensation of the unstable mode alone without relaxation (with the energy difference taken at the maximum gain, i.e., the depth of the double well). We can see that energy gain is small for LDA, large for PBE and intermediate for PBEsol functionals, as usually found in ferroelectric crystals. Also, it can be noticed that SOC enhances energy gain in all computed cases.}
\end{center}
\end{table}

The analytical model used in the main text to fit the DFT energies as a function of the invariant and polar distortions amplitudes ($Q_{1}$ and $Q_{2}$ respectively) is also reported here:


\begin{align}\label{Eq:Umodel_supp}
U(Q_1,Q_2) &\ = \frac{\alpha_1}{2}  Q_1^2 + \frac{\alpha_2}{2} Q_2^2+\frac{\beta_2}{4} Q_2^4  + aQ_{1}Q_{2}^{2} + V_{\text{high}}(Q_1,Q_2)
\end{align}   

with

\begin{equation}
    V_{\text{high}}(Q_1,Q_2) = \frac{\beta_1}{3}Q_1^3 + \frac{\gamma_2}{6}Q_2^6 + bQ_{1}^{2}Q_{2}^2 + cQ_1Q_2^4.
\end{equation}

The usual mean field approximation $\alpha' = \alpha'_{0}(T-T_c)$ ($\alpha'_{0}>0$) is assumed. $Q_1$ and $Q_2$ are scaled by the amplitudes of the $\Gamma_1$ and $\Gamma_2$ SAMs projection of the ferro-distortion as reported in tab~\ref{tab:Ampli_SAMs}. 

We have adopted the following fitting strategy: given the high and low symmetry phases, with the help of AMPLIMODES we have extracted the distorted structures corresponding to the symmetry-adapted modes ($\Gamma_1$ and $\Gamma_2$ IRs) at various amplitudes ranging from 0 to 2 (0 represents the paraelectric reference while 1 is referred to the SAM amplitude associated with the ground state). First, we have restricted the interpolation process to small  $Q_1$ and $Q_2$ to make the contribution of V$_\text{high}$ negligible. It may also be noted that if we consider the low-energy part of the model only, taking $\partial{U}/\partial{Q_1} = 0$ leads to a quartic expression in term of the polar mode only, that is $U_{\text{eff}}(Q_2) = \frac{\alpha_2}{2} Q_2^2+\frac{\beta'_2}{4} Q_2^4$ with $\beta'_2 \equiv \beta_2 - 2a^{2}/\alpha_1$. The sign of $\beta'_2$ is important since it is directly associated with the order of the phase transition. Furthermore, with the knowledge of the $\Gamma_{2}$ equilibrium amplitudes and energies of either the soft mode only ($\Delta{E}_\text{soft} = 13.0$ meV and A$_{\Gamma_{2};\text{soft}} = 0.6177$ \AA) and of the full distortion ($\Delta{E}_\text{tot.} = 89.0$ meV and A$_{\Gamma_{2};\text{tot.}} = 1.4670$ \AA), we can extract $\alpha_2^{in.} \equiv -4|\Delta{E}_\text{soft}|/$/A$_{\Gamma_{2};\text{soft}}^2 =$  136.3 meV/\AA$^{2}$ and $\beta_2^{in.} \equiv 4|\Delta{E}_\text{soft}|/$/A$_{\Gamma_{2};\text{soft}}^4 =$ 357.2 meV/\AA$^{4}$, which can be used to initialise the fitting process. Similarly, $a$ can be initialised from the relations $\beta'_2 = 4|\Delta{E}_\text{tot.}|/$/A$_{\Gamma_{2};\text{tot.}}^4$  and $a^{in.} = -\sqrt{\alpha_1(\beta_2-\beta'_2)}$, leaving in principle only $\alpha_1$ as a "free parameter". Once the low-energy model ($U(Q_1,Q_2) - V_{\text{high}}(Q_1,Q_2)$) has been determined, we have extended the interpolation to V$_{\text{high}}$ to improve the modeling of the large-amplitude behavior. 

\begin{figure}
     \includegraphics[width=0.6\textwidth]{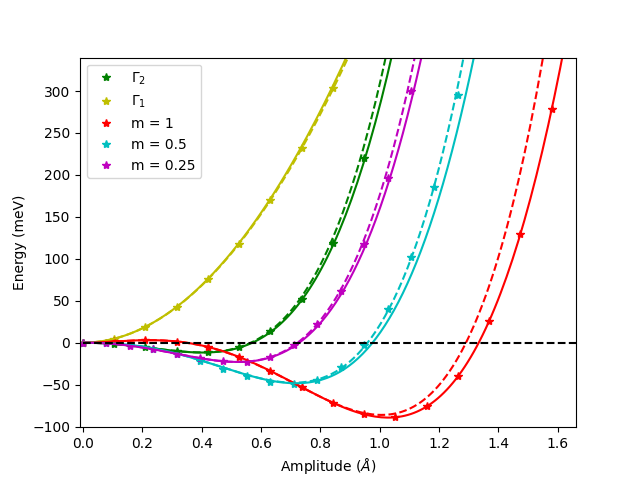}
     \caption{Energy landscape of PGO as a function of the amplitude of the SAM $\Gamma_1$ and $\Gamma_2$. Symbols are SOC-including DFT calculations while the lines correspond to the model of eq.~\ref{Eq:Umodel_supp}, either with (full) and without (dashed) V$_{\text{high}}$ contribution. The value of $m$ in the caption multiplies the $A_{\Gamma_{1}}/A_{\Gamma_{2}} = 0.218$ ratio associated with the linear distortion from the $P\Bar{6}$ to the $P3$ ground states, as extracted from AMPLIMODES.}
     \label{fig:test_landscape}
     \end{figure}

We report now the values of the coefficients (f.u.): $\alpha_1 = 6793.7$ meV/\AA$^{2}$, $\beta_1 = 634.0$ meV/\AA$^{3}$, $\alpha_2 = -135.0$ meV/\AA$^{2}$, $\beta_2 = 391.7$ meV/\AA$^{4}$, $\gamma_2 = -15.6$ meV/\AA$^{6}$, $a = -1077.6$ meV/\AA$^{3}$, $b=-114.0$ meV/\AA$^{4}$, $c=28.0$ meV/\AA$^{6}$ and $\beta'_2=$ 49.8 meV/\AA$^{4}$. From the inspection of the model parameters, we realise the following facts. First, spin-orbit coupling and lattice optimization have a substantial impact. However, most of the PE-FE energy barrier is due to internal ion relaxation, as can also be inspected from tab.~\ref{tab:DeltaE}. Secondly, the quartic (positive) polar coefficient $\beta_2$ is strongly renormalized by the interaction with the invariant mode, even though it maintains its sign, meaning that the second-order character of the phase transition observed in the experiments is correctly reproduced by our calculations. Moreover, even if the $Q_2^{6}$ polar coefficient appears harmful, it is counterbalanced by the high-order favorable terms of the interaction, which ensure the stability of the ferroelectric solution. We have tested~\ref{fig:test_landscape} our model on $A_{\Gamma_{1}}/A_{\Gamma_{2}}$ ratios other than the PE-FE transition, finding an adequate level of transferability.

\subsection{Polarisation in open circuit boundary conditions}

As we discuss in the main section, a polar soft LO branch is present, therefore we investigate the possibility that PGO may be present a nonzero spontaneous polarisation $\mathbf{P}_s$ in open circuit boundary conditions (OCBC), which corresponds to the case where $\mathbf{D} = \epsilon\mathbf{E} + \mathbf{P_s}$ = 0. Following~\cite{PhysRevB.107.094108,PhysRevB.99.104101}, we may write the electric free energy as:

\begin{equation}\label{eq:elec_model}
 F(Q_1,Q_2,E) = U(Q_1,Q_2) + \Omega(Q_1,Q_2)\frac{1+\chi_{\infty}(Q_1,Q_2)/2}{\epsilon_0[1+\chi_{\infty}(Q_1,Q_2)]^2}P_s^2(Q_2),
\end{equation}

where $\Omega$ is the volume and $\chi_{\infty}$ is the high-frequency dielectric susceptibility. Eq.~\ref{eq:elec_model} generalises eq.(1) of ref.~\cite{PhysRevB.107.094108} to multiple distortions and may be used to probe the energy landscape in OPBC. However and due to the size of PGO, fitting such equation would be too computationally demanding, hence sensible approximations are required. One can safely assume a linear relation $P_s(Q_2) = pQ_2$ such that, for instance, the equilibrium $Q_2$ returns the spontaneous polarisation computed with the Berry phase. Likewise, the volume is also poorly affected by the relaxation: $\Omega(P\Bar{6}) =$ 1030 \AA$^3$ and $\Omega(P3)$ = 1040 \AA$^3$ (SOC included) so that a linear interpolation across the phase transition between these two very close values or even take a fixed value (e.g. the average 1035 \AA$^{3}$)is sufficient. The susceptibility may change nontrivially as a function of the two SAMs~\cite{PhysRevB.107.094108,PhysRevB.99.104101}, although $\chi_{\infty}(P\Bar{6}) - \chi_{\infty}(P3)$ (probed at the ground state phases) is also expected to be small. A further simplification comes from dropping the $Q_1$ dependency on both $\chi^{\infty}$ and $\Omega$. With a dielectric permittivity of 25 $\epsilon_0$~\cite{conroy2023observation}, we may approximate eq.~\ref{eq:elec_model} to:

\begin{equation}\label{eq:elec_model_approx}
 F(Q_1,Q_2,E) \sim U(Q_1,Q_2) + \Omega(Q_2)\frac{P_s^2(Q_2)}{2\epsilon(Q_2)}.
\end{equation}

If we linearise $\Omega(x) = \Omega(P\Bar{6}) + \Delta\Omega{x}$, $\epsilon(x) = \epsilon(P\Bar{6}) + \Delta\epsilon{x}$ we can easily prove that only even terms in the polar amplitude x are allowed by symmetry. Thus, $\Omega(Q_2)P_s^2(Q_2)/\epsilon(Q_2) \sim (\Omega(P\Bar{6})p^2/\epsilon(P\Bar{6}))Q_2^2 +(\Delta\Omega\cdot\Delta\epsilon)({p^2}/\epsilon(P\Bar{6})^2)Q_2^4 + O(Q_2^6)$. We hence expect the quartic correction to be rather small, so that the OPBC physics is dominated by the second order term, which hardens the $\alpha_2$ parameter:

\begin{equation}
     \alpha_2(\mathbf{D}=0) = \alpha_2 + \frac{P_s^2}{\epsilon},
\end{equation}
   
where P$_{s}$ = 2.8 $\mu$C/cm$^{2}$ is the Berry phase calculated in Q$_{1}$ = 0 condition. Since $\alpha_2$ = -135.0 meV/Angstrom$^{2}$ and $P_s^2/\epsilon$ = 15.5 meV/Angstrom$^{2}$, the ferroelectricity is in fact preserved. From the second-order correction alone we can estimate $P(\mathbf{D}=0) \sim P_s \sqrt{\alpha_2(\mathbf{D}=0)/\alpha_2}$ $\sim$ (2.6 $\mu$C$\cdot$cm$^{-2}$) 5.2 $\mu$C$\cdot$cm$^{-2}$ with the invariant mode (excluded) included. Note that the calculation from the phonon spectrum (e.g. the TO vs LO mode) gives a lower value of 4.0 $\mu$C$\cdot$cm$^{-2}$ (reported in the main text), most likely because of the differences between phonon calculations and the fitting of the energy landscape at small amplitudes. Nevertheless, both approaches support a finite polarisation in open circuit boundary conditions and thus hyperferroelectricity in the PGO compound.

\end{document}